\documentclass[%
 reprint,
 amsmath,amssymb,
 aps,
prb,
]{revtex4-1}

\usepackage{graphicx}
\usepackage{dcolumn}
\usepackage{bm}
\usepackage{color}

\begin{document}

\preprint{APS/123-QED}

\title{Collinear Spin-density-wave Order and Anisotropic Spin Fluctuations \\
in the Frustrated $J_1$--$J_2$ Chain Magnet $\mathrm{NaCuMoO_4(OH)}$}
\author{Kazuhiro~Nawa}
\altaffiliation{Present address: Institute of Multidisciplinary Research for Advanced Materials, Tohoku University, 2-1-1 Katahira, Sendai 980-8577, Japan}
\email{knawa@tagen.tohoku.ac.jp}
\author{Makoto~Yoshida}
\altaffiliation{Present address: Max-Planck Institute for Solid State Research, Stuttgart, 70569 Stuttgart, Germany}
\author{Masashi~Takigawa}
\author{Yoshihiko~Okamoto}
\altaffiliation{Present address: Department of Applied Physics, Nagoya University, Nagoya 464-8603, Japan}
\author{Zenji~Hiroi}
\affiliation{%
Institute for Solid State Physics, The University of Tokyo, Kashiwa, Chiba 277-8581, Japan
}%

\date{\today}

\begin{abstract}
The phase diagram of the quasi-one-dimensional magnet $\mathrm{NaCuMoO_4(OH)}$ is established through
single-crystal NMR and heat-capacity measurements.
The $^{23}$Na and $^1$H NMR experiments indicate a spiral and a collinear
spin-density-wave (SDW) order below and above $B_c$ = 1.5-1.8 T, respectively.
Moreover, in the paramagnetic state above the SDW transition temperature, the nuclear spin-lattice relaxation rate
$1/T_1$ indicates anisotropic spin fluctuations that have gapped excitations in the transverse spectrum
but gapless ones in the longitudinal spectrum. These static and dynamic properties are well
described by a theoretical model assuming quasi-one-dimensional chains with competing ferromagnetic
nearest-neighbor interactions $J_1$ and antiferromagnetic next-nearest-neighbor interactions $J_2$ ($J_1$--$J_2$ chains).
Because of the excellent crystal quality and good one dimensionality, 
$\mathrm{NaCuMoO_4(OH)}$ is a promising compound to elucidate the unique physics of the frustrated $J_1$--$J_2$ chain.
\end{abstract}

\pacs{Valid PACS appear here}
\maketitle


\section{\label{introduction}Introduction} 
Frustrated magnets with competing magnetic interactions are expected to exhibit exotic ground states
such as spin liquids \cite{SL, SL2, SL3}, valence bond solids \cite{SL2, VBC},
and spin nematic states \cite{nematic, nematic2, nematic3, 1Dtheory00, 1Dtheory0, 1Dtheory1, 1Dtheory2, 1Dtheory3, 1Dtheory4, 1Dtheory5, 1Dtheory6}.
Among them, the one-dimensional (1D) spin-1/2 system with ferromagnetic nearest-neighbor interactions $J_1$ frustrating
with antiferromagnetic next-nearest-neighbor interactions $J_2$ 
has recently drawn much attention, because this model exhibits rich quantum phases in magnetic fields $B$ 
\cite{1Dtheory00, 1Dtheory0, 1Dtheory1, 1Dtheory2, 1Dtheory3, 1Dtheory4, 1Dtheory5, 1Dtheory6, 1Dtheory7}.
Particularly interesting is a spin nematic state expected near the fully polarlized state 
\cite{1Dtheory00, 1Dtheory0, 1Dtheory1, 1Dtheory2, 1Dtheory3, 1Dtheory4, 1Dtheory5, 1Dtheory6, 1Dtheory7}.
In an ordinary magnet, when the magnetic field is decreased below the saturation field, a conventional magnetic order sets in as a result of 
the Bose-Einstein condensation of single magnons. In contrast, in a quasi-1D frustrated $J_1$--$J_2$ chain model,
the Bose-Einstein condensation of bound magnons leads to a spin-nematic order, where rotation symmetry perpendicular to the magnetic field is
broken while time-reversal symmetry is preserved. When the magnetic field is further decreased, bound magnons form a spin-density-wave (SDW) order.
Near zero field, bound magnons are destabilized and a spiral order occurs.

Experimental studies have revealed that several materials reflect the quasi-1D frustrated $J_1$--$J_2$ chain model,
such as $\mathrm{LiCuVO_4}$\cite{neutron00, neutron0, NMR1, PD_LCVO, neutron1, neutron2, NMR3, NMR4, NMR2, NMR5, magnetization, HFNMR, HFHC, HFNMR2},
$\mathrm{Li_2ZrCuO_4}$\cite{Li2ZrCuO4_0, Li2ZrCuO4},
$\mathrm{Rb_2Cu_2Mo_3O_{12}}$\cite{Rb2Cu2Mo3O12_0, Rb2Cu2Mo3O12},
$\mathrm{PbCu(SO_4)(OH)_2}$\cite{PbCuSO4OH_0, PbCuSO4OH_2, PbCuSO4OH_3, PbCuSO4OH_4, PbCuSO4OH_6, PbCuSO4OH_7, PbCuSO4OH_8, PbCuSO4OH_9},
$\mathrm{LiCuSbO_4}$\cite{LiCuSbO4, LiCuSbO4_2},
$\mathrm{LiCu_2O_2}$\cite{LiCu2O2, LiCu2O2_neu, LiCu2O2_3},
3-I-V\cite{3IV},
and $\mathrm{TeVO_4}$\cite{TeVO4, TeVO4_2, TeVO4_3, TeVO4_4}.
Among them,
$\mathrm{LiCuVO_4}$ has been most extensively studied.
It exhibits an incommensurate spiral order at low fields and an incommensurate SDW order at intermediate fields above 7 T
\cite{NMR1, PD_LCVO, neutron1, neutron2, NMR3, NMR4, NMR2}.
In addition, recent NMR experiments revealed the coexistence of gapped transverse excitations and
gapless longitudinal excitations above the transition temperature of the SDW order, which indicates the formation of bound magnon pairs \cite{NMR2, NMR5}.
The linear field dependence of magnetization observed between 40.5 and 44.4~T was initially interpreted as a signature of a spin nematic order\cite{magnetization}.
However, its origin remains debated, since further NMR studies in steady magnetic fields revealed that the field dependence of the NMR internal field is different from the magnetization curve\cite{HFNMR}.
The internal field becomes constant above 41.4 T at 0.38~K, indicating that the linear variation of the magnetization is due to inhomogeneity induced by Li deficiency\cite{HFNMR}.
On the other hand, recent NMR experiments in pulsed magnetic fields at 1.3~K indicate that the internal field exhibits a linear variation between 42.41 and 43.55~T without inhomogeneity\cite{HFNMR2}.
The origin of the discrepancy between the two NMR results is unclear at present. It might be related to differences in sample quality or measured temperature.
Since the broad $^{51}$V NMR spectra in $\mathrm{LiCuVO_4}$ make it difficult to obtain direct evidence of the spin nematic state, a new candidate having less crystalline defects is greatly desired.

Recently, $\mathrm{NaCuMoO_4(OH)}$ was proposed as a candidate $J_1$--$J_2$ chain magnet \cite{NaCuMoO4OH, NaCuMoO4OH3}.
It crystallizes in an orthorhombic structure with the space group $Pnma$ and consists of edge-sharing $\mathrm{CuO_4}$ plaquettes, which form $S$ = 1/2 chains along the $b$ axis, as shown in Fig.~\ref{cryst}(a).\cite{NaCuMoO4OH2}.
From the magnetization, $J_1$ and $J_2$ are estimated as $-$51~K and 36~K, respectively \cite{NaCuMoO4OH}.
The magnetic order is observed below $T_\mathrm{N}$ =~0.6~K at zero field \cite{NaCuMoO4OH},
which is lower than 2.1~K for $\mathrm{LiCuVO_4}$ \cite{neutron00}, indicating a good 1D character.
In addition, the saturation field of 26~T is lower than the value of 41~T for $\mathrm{LiCuVO_4}$ \cite{magnetization, HFNMR},
which is greatly advantageous for experiments, especially to explore the spin nematic phase immediately below the saturation field.
However, the features of magnetic ground states and spin fluctuations have not yet been determined because of the lack of a single crystal.

In this paper, we report NMR and heat-capacity measurements on a single crystal of $\mathrm{NaCuMoO_4(OH)}$.
The remainder of this paper is organized as follows.
The experimental setup for NMR and heat-capacity measurements is described in Section \ref{exp}.
Their results are presented in Section \ref{rad}.
First, the coupling tensor is estimated from $K-\chi$ plots in Section~\ref{para},
and then the phase diagram is established from NMR and heat-capacity measurements in Section~\ref{PD}.
In Section~\ref{order}, NMR spectra in ordered phases are shown and compared with simulated curves.
The NMR spectra indicate the occurrence of an incommensurate spiral order below a transition field $B_c$ of 1.5--1.8~T and a collinear SDW order above $B_c$.
In Section~\ref{T1subsec}, spin fluctuations are discussed from a spin-relaxation rate $1/T_1$ .
The temperature dependence of $1/T_1$ above $B_c$ indicates the development of anisotropic spin fluctuations with gapped transverse excitations above the SDW transition temperature.
In Section~\ref{dis}, the magnetic properties of $\mathrm{NaCuMoO_4(OH)}$ are compared with those of other candidates.
For instance, disorder effects due to crystalline defects are smaller than those in $\mathrm{LiCuVO_4}$, indicating
$\mathrm{NaCuMoO_4(OH)}$ is a more ideal compound for studying the frustrated $J_1$--$J_2$ chain.
Finally, a summary is presented in Section~\ref{summary}.

\section{\label{exp}Experiments}
\begin{figure}[t]
\includegraphics[width=8.5cm]{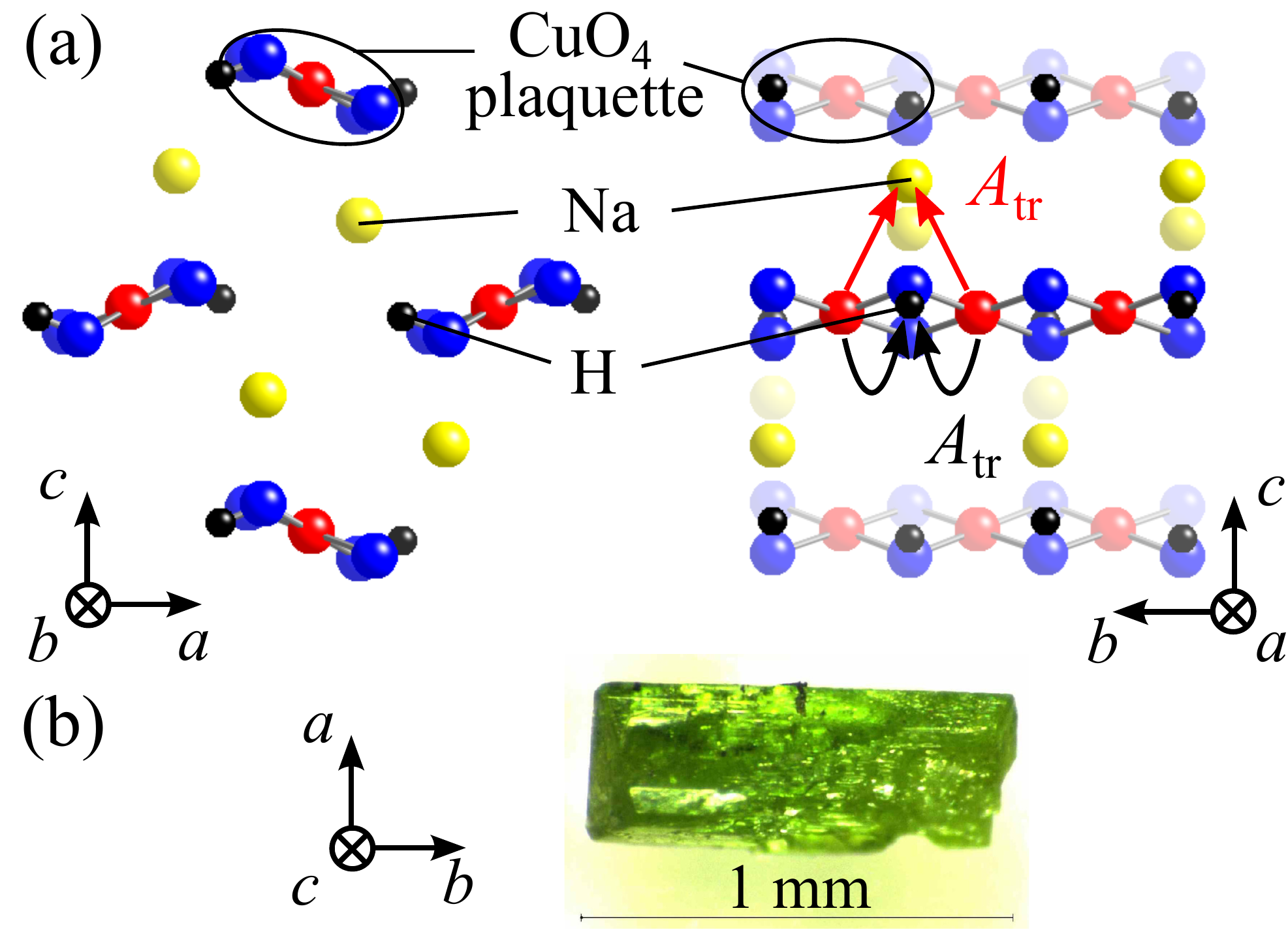}
\caption{\label{cryst}(a) Crystal structure of $\mathrm{NaCuMoO_4(OH)}$ along $b$- and $a$-axes.
Red, yellow, black, and blues spheres represent Cu, Na, H, and O atoms, respectively,
while Mo and a part of O atoms are omitted for clarity.
Arrows schematically describe transferred hyperfine interactions of Na and H nuclei with their two nearest Cu sites.
(b) Photograph of the single-crystalline $\mathrm{NaCuMoO_4(OH)}$ used for heat-capacity and NMR measurements.}
\end{figure}

We used a single crystal grown by a hydrothermal method \cite{NaCuMoO4OH3} with a size of 0.4$\times$0.4$\times$1.0~$\mathrm{mm}^3$, a photograph of which is shown in Fig.~\ref{cryst}(b).
Although the crystal includes a small amount of lindgrenite (less than 1\% in a molar mass), its influence is negligible
since NMR spectra and $1/T_1$ do not show a visible change even at its ferrimagnetic transition temperature of 14 K.
NMR experiments were performed on $^{23}$Na ($^{23}\gamma$/($2 \pi$) = 11.26226 MHz/T, $I$ = 3/2) and $^1$H ($^1\gamma$/($2 \pi$) = 42.57639 MHz/T, $I$ = 1/2) nuclei.
A two-axis piezo-rotator combined with a dilution refrigerator enables our NMR experiments below 1 K with
small misorientations within 3$^\circ$ for $B \parallel c$ and 5$^\circ$ for $B \parallel a$.
The NMR spectra were obtained by summing the Fourier transform of the spin-echo signals obtained at equally spaced rf frequencies.
$1/T_1$ was determined by the inversion recovery method.
The time evolution of the spin-echo intensity for $^{23}$Na and $^1$H nuclei was fitted to a theoretical recovery curve 
of $M(t) = M_{eq} - M_0 [ 0.1\exp\{-(t/T_1)^\beta\}+0.9\exp\{-(6t/T_1)^\beta\} ]$\cite{relaxation, relaxation2} 
and $M(t) = M_{eq} - M_0\exp\{-(t/T_1)^\beta\}$, respectively, where
$\beta$ is a stretch exponent indicating the distribution of $1/T_1$.
It becomes smaller than 1 because of an incommensurate magnetic order below $T_N$, while it is fixed to 1 above $T_N$.
Heat capacity was measured by the relaxation method (PPMS, Quantum Design).
The magnetic heat capacity is obtained by subtracting the phonon contribution, which is estimated from a Zn-analogue\cite{NaCuMoO4OH}.

.
\section{\label{rad}Results and Discussions}
\subsection{\label{para}Estimation of coupling tensors} 
First, we discuss hyperfine coupling tensors $\mathbf{A}$ for both $^{23}$Na and $^{1}$H nuclei,
which are necessary to determine magnetic structures from NMR spectra (see Section~\ref{order})
and spin fluctuations from $1/T_1$ quantitatively (see Section~\ref{T1subsec}).
The positions of Na and H atoms in $\mathrm{NaCuMoO_4(OH)}$ are illustrated in Fig.~\ref{cryst}(a).
Na atoms are located in the middle of two Cu chains consisting of edge-sharing $\mathrm{CuO_4}$ plaquettes,
and H atoms are bonded to O atoms on $\mathrm{CuO_4}$ plaquettes.
All Na or H atoms are crystallographically equivalent and occupy $4c$ sites.
They are also symmetrically equivalent for $B \parallel ab$ and $B \parallel bc$ 
while they can split into two inequivalent sites when the magnetic field is applied along the other directions.
The atomic coordinates of Na and H atoms are (0.3697(5), 1/4, 0.3056(4))\cite{NaCuMoO4OH2} and (0.243(4), 1/4, 0.030(4))\cite{NaCuMoO4OH4}, respectively.
The atomic position of H atoms was determined from neutron diffraction experiments\cite{NaCuMoO4OH4}, and it was also confirmed by density functional theory (DFT) calculations with the generalized gradient approximation plus onsite repulsion $U$,
which yield  (0.25015, 1/4, 0.01952). The detailed procedure of DFT calculations is the same as described in Ref.~\onlinecite{DFT}.

The internal field $\mathbf{B}_\mathrm{int}$ at a ligand nucleus is expressed by $\mathbf{B}_\mathrm{int} = \sum_i \mathbf{A}^i \cdot \mathbf{\mu}^i$,
where $\mathbf{A}^i$ is the hyperfine coupling tensor and $\mathbf{\mu}^i$ is the magnetic moment of the $i$-th Cu site.
$\sum_i \mathbf{A}^i$ appears in a linear relation between the magnetic shift $\mathbf{K}$
and magnetic susceptibility $\chi$ in the paramagnetic phase:
\begin{equation}
\mathbf{K} = \frac{1}{N \mu_B} \sum_i \mathbf{A}^i \cdot \mathbf{\chi}. \label{linear}
\end{equation}
We first determined $\sum_i \mathbf{A}^i$ from Eq.~\eqref{linear} experimentally and then estimated each $\mathbf{A}^i$.

\begin{figure}[t]
\includegraphics[width=8.5cm]{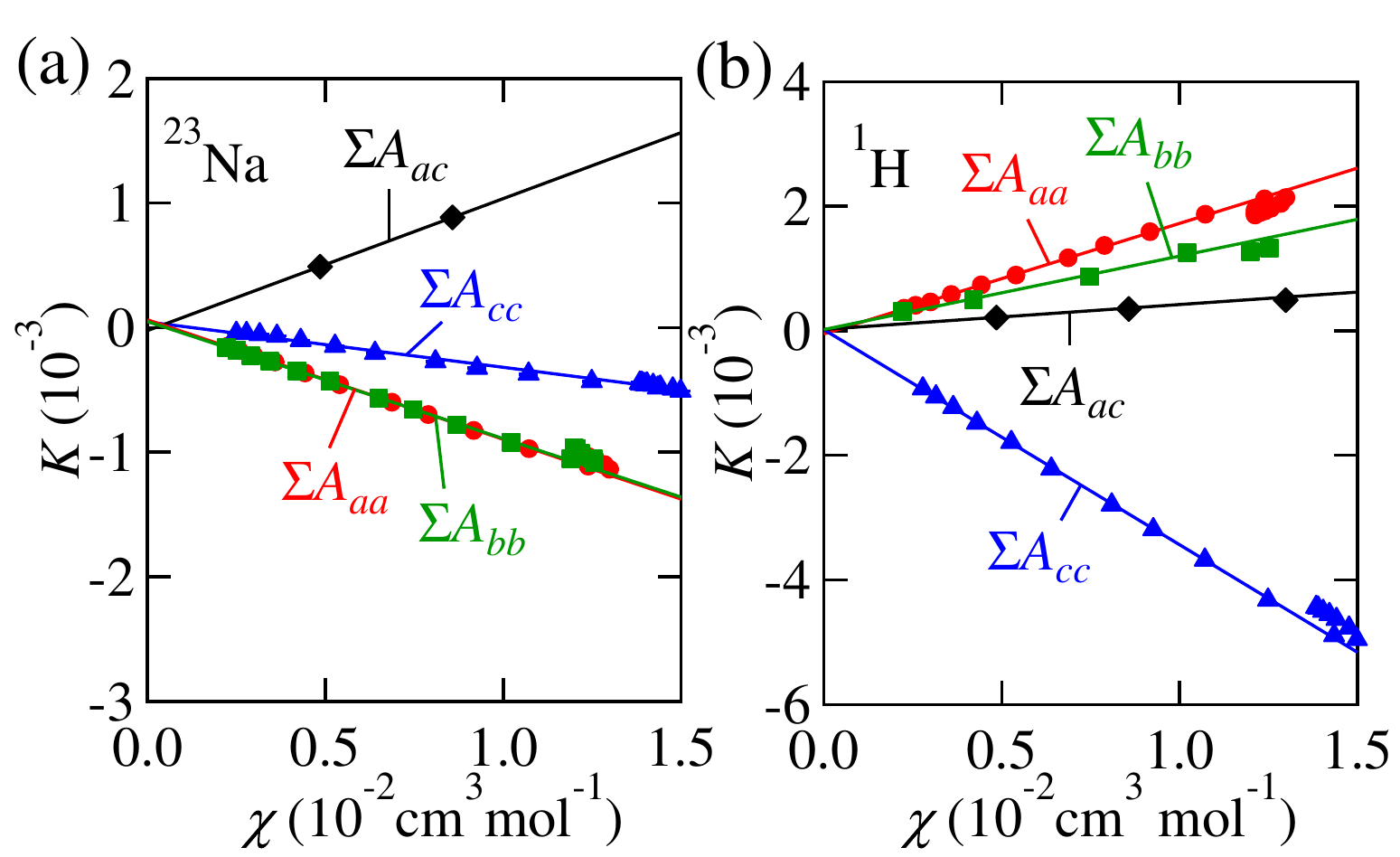}
\caption{\label{Kchi} $K$--$\chi$ plots for (a) $^{23}$Na and (b) $^{1}$H nuclei.
Lines indicate linear fits to estimate coupling tensors, which are listed in Table~\ref{hyperfine}.}
\end{figure}

The linear relation \eqref{linear} for the diagonal components $K_{\epsilon \epsilon} (\epsilon = a, b, c)$
are confirmed by the $K$--$\chi$ plots shown in Fig.~\ref{Kchi}.
They are defined by the observed resonance frequency 
\begin{equation}
\nu_\mathrm{res}^{(\epsilon )} = (1 + K_{\epsilon \epsilon }) \gamma B \label{nu1},
\end{equation}
and $\sum_i A_{\epsilon \epsilon}^i$ is determined from the linear slope of the $K_{\epsilon \epsilon}$--$\chi$ plot.
The values of $\sum_i A_{\epsilon \epsilon}^i$ determined experimentally are listed as $\sum_i \mathbf{A}_\mathrm{exp}^i$ in Table.~\ref{hyperfine}.

\begin{figure}[t]
\includegraphics[width=8.5cm]{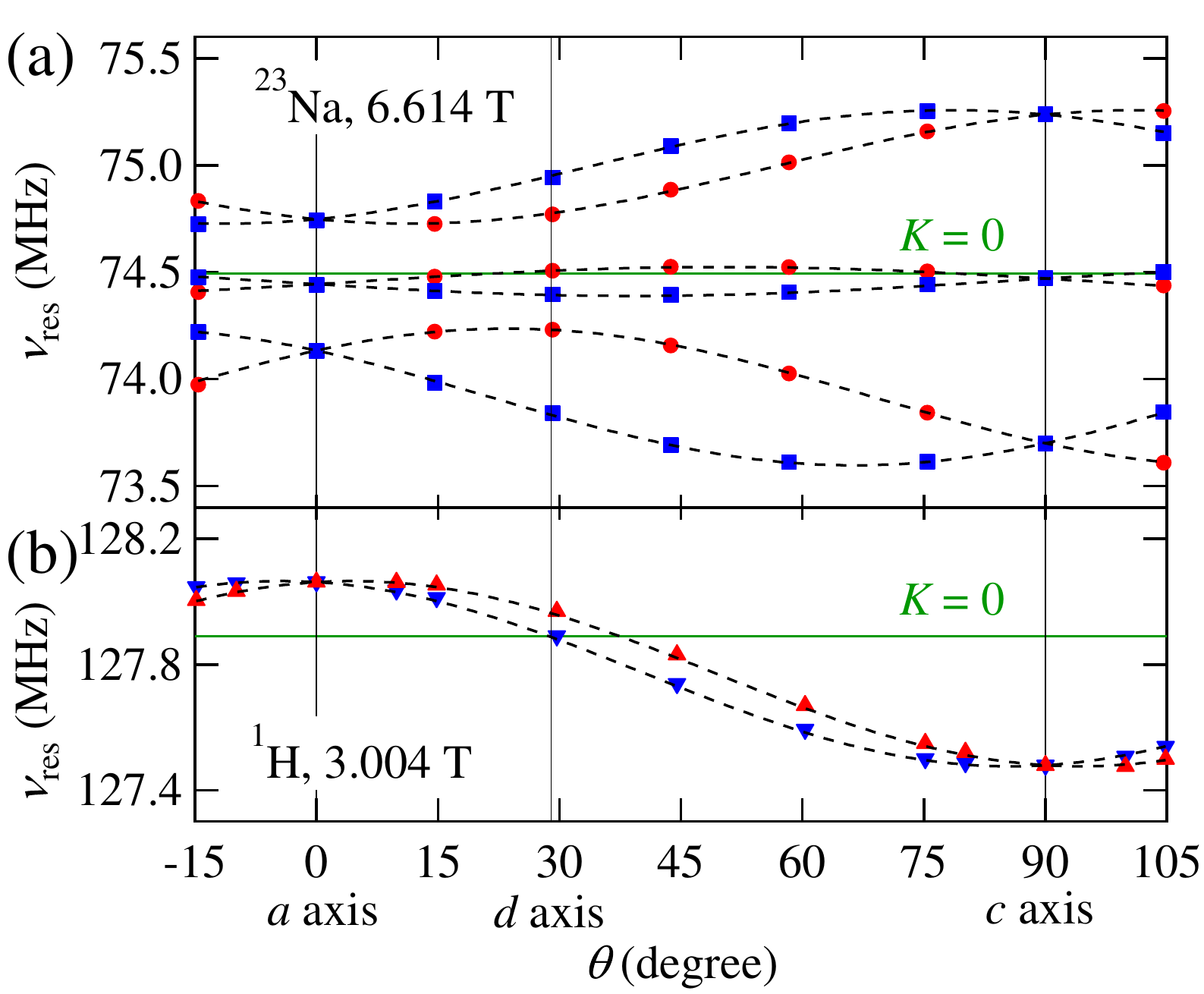}
\caption{\label{rotation} Angle dependences of resonance frequencies for (a) $^{23}$Na and (b) $^1$H nuclei at 50 K.
Two inequivalent sites are colored in red and blue. Dashed curves represent the fitting curves of Eq.~\eqref{nu2} for $^1$H
and those of Eq.~\eqref{nu3} for $^{23}$Na.
Horizontal lines represent frequencies corresponding to $K=0$. The $d$-axis is the specific axis defined in the main text.}
\end{figure}

The nondiagonal components also follow the linear relation \eqref{linear}.
While $K_{ab}$ and $K_{bc}$ become 0 because of symmetry and, thus, $A_{ab}$ and $A_{bc}$ cannot be determined,
$K_{ac}$ can be determined by the angle dependences of the resonance frequency $\mathbf{\nu}_\mathrm{res}$.
For $B \parallel ac$, the crystallographically equivalent sites can split into two inequivalent sites for either $^{23}$Na or $^1$H.
In fact, two resonance lines are observed in the $^1$H NMR spectra.
Their angle dependences $\mathbf{\nu}(\theta)_\mathrm{res}$ are fitted to the following function:
\begin{equation}
\begin{split}
\nu(\theta)_\mathrm{res} &= (1 + K_{aa} \cos^2 \theta \pm 2 K_{ac} \sin \theta \cos \theta \\
&+ K_{cc} \sin^2 \theta) \gamma B,  \label{nu2}
\end{split}
\end{equation}
with $K_{aa}$ and $K_{cc}$ fixed to the values determined from Eq.~\eqref{nu1}.
This fit reproduces $\mathbf{\nu}(\theta)_\mathrm{res}$ well, as shown in Fig.~\ref{rotation}(b),
and yields $K_{ac} = 3.51 \times 10^{-3}$ by using $K_{aa} = 1.37 \times 10^{-3}$ and $K_{cc} = -3.19 \times 10^{-3}$ at 50 K.

For $^{23}$Na nuclei, the quadrupole interaction produces three peaks per site.
Thus, six resonance lines are observed in $^{23}$Na spectra.
The angle dependence of their positions are fit to the functions
including the contributions of the magnetic shift and quadrupole splitting \cite{angle}:
\begin{align}
&\nu^I_{m, m-1}(\theta)_\mathrm{res} = (1 + K_{aa} \cos^2 \theta \pm 2 K_{ac} \sin \theta \cos \theta \notag \\
&+ K_{cc} \sin^2 \theta) \gamma B \notag \\
&- \frac{1}{2} \left( m - \frac{1}{2} \right) \nu_Q + \frac{1}{2} \left( m - \frac{1}{2} \right) \nu_Q \eta \cos 2 (\theta - \theta_Q) \notag \\
&-\frac{\nu_Q^2}{32\gamma B} \{ 6m(m-1) - 2I(I+1)+3 \} \notag \\
& \ \ \ \times \left( 1 + \frac{2}{3} \eta \cos 2(\theta \mp \theta _Q) \right) \label{nu3} \\
&+\frac{\nu_Q^2 \eta^2}{72 \gamma B} \Bigg[ 24m(m-1) - 4I(I+1) + 9 \notag \\
&-\left\{ \frac{51}{2} m(m-1) - \frac{9}{2} I(I+1) + \frac{39}{4} \right\} \notag \\
& \ \ \ \times \cos^2 2(\theta \mp \theta_Q) \Bigg]. \notag 
\end{align}
$I$ and $m$ are constants that represent a nuclear spin of 3/2 and its $z$-component (3/2, 1/2, or -1/2), respectively.
$\nu_Q$ is a quadruplole frequency along the maximam principal axis,
$\eta$ is an asymmetry parameter,
and $\theta_Q$ is the angle between the $a$-axis and the closest principal axis of the electric-field gradient; 
note that the principal axes of the electric-field gradient exist in the $ac$-plane and along the $b$-axis. The
free parameters in this fit are $\eta$, $\theta_Q$, and $K_{ac}$.
$\nu_Q$ is determined from NMR spectra for $B~\parallel~b$, and
$K_{aa}$ and $K_{cc}$ are fixed at the values determined from Eq.~\eqref{nu1}.
The fit at 50~K, which is shown in Fig.~\ref{rotation}(a), reproduces $\mathbf{\nu}(\theta)_\mathrm{res}$ well and
yields $\eta$ = 0.532, $\theta_Q$ = 18.1$^\circ$, and $K_{ac} = 8.85 \times 10^{-4}$
by using $\nu_Q$ = 1.074 MHz, $K_{aa} = -6.98 \times 10^{-4}$, and $K_{cc} = -3.24 \times 10^{-4}$.
We determined $\sum_i A_{ac}^i$ from the linear slope of the $K_{ac}$--$\chi$ plot, as shown in Fig.~\ref{Kchi}.
Their values are listed in Table~\ref{hyperfine}.

\begin{table}[t]
\caption{Coupling tensors for $^{23}$Na and $^1$H nuclei.
$\sum_i \mathbf{A}^i_\mathrm{exp}$, $\sum_i \mathbf{A}^i_\mathrm{dip}$, and $\sum_i \mathbf{A}^i_\mathrm{tr}$ describe
the total, dipolar, and transferred hyperfine contributions, respectively.
$\sum_i \mathbf{A}^i_\mathrm{dip}$ includes the sum of the Lorentz and demagnetization field,
which is estimated as $-$0.010 ($aa$), 0.020 ($bb$), $-$0.010~T/$\mu_\mathrm{B}$ ($cc$) \cite{demag}.
The values of the transferred hyperfine coupling adopted for the simulation are listed as $\sum_i \mathbf{A}^i_\mathrm{tr, sim}$. All values are described in units of T/$\mu_\mathrm{B}$.}
\label{hyperfine}
\begin{center}
\begin{tabular}{cccccc}
\hline
 & & $\sum_i \mathbf{A}^i_\mathrm{exp}$ & $\sum_i \mathbf{A}^i_\mathrm{dip}$
& $\sum_i \mathbf{A}^i_\mathrm{tr}$ & $\sum_i \mathbf{A}^i_\mathrm{tr, sim}$ \\
\hline
& $aa$ & $-$0.054(10) & $-$0.015(10) & $-$0.039(14) & $-$0.050 \\
$^{23}$Na & $bb$ & $-$0.052(10) & $-$0.009(10) & $-$0.043(14) & $-$0.043 \\
& $cc$ & $-$0.022(10) & 0.024(10) & $-$0.046(14) & $-$0.038 \\
& $ac$ & 0.059(10) & 0.066(10) & $-$0.007(14) & 0 \\
\hline
 & $aa$ & 0.099(10) & 0.085(10) & 0.014(14) & 0.014 \\
$^{1}$H & $bb$ & 0.066(10) & 0.031(10) & 0.035(14) & 0.035 \\
 & $cc$ & $-$0.193(10) & $-$0.115(10) & $-$0.078(14) & $-$0.078 \\
& $ac$ & 0.021(10) & 0.036(10) & $-$0.015(14) & $-$0.010 \\
\hline
\end{tabular}
\end{center}
\end{table}

Next, we estimated $\mathbf{A}^i$ from the coupling tensor determined experimentally, $\sum_i \mathbf{A}_\mathrm{exp}^i$, in the following manner.
$\sum_i \mathbf{A}_\mathrm{exp}^i$ can be divided into two contributions: $\sum_i \mathbf{A}_\mathrm{dip}^i$ and $\sum_i \mathbf{A}_\mathrm{tr}^i$.
$\sum_i \mathbf{A}_\mathrm{dip}^i$ is calculated by a lattice sum of dipolar interactions within a sphere with a radius of 60~\AA \,
together with a Lorentz field and a demagnetization field.
The sum of the Lorentz and demagnetization field is estimated as $-$0.010, 0.020, $-$0.010~T/$\mu_\mathrm{B}$ for the $a$-, $b$-, $c$-components, respectively, from the crystal shape\cite{demag}.
The contribution of transferred hyperfine interactions corresponds to the difference,
$\sum_i \mathbf{A}_\mathrm{tr}^i \equiv \sum_i \mathbf{A}_\mathrm{exp}^i - \sum_i \mathbf{A}_\mathrm{dip}^i$.
We assumed that $\sum_i \mathbf{A}^i_\mathrm{tr}$ consists of contributions from only two nearest-neighbor Cu sites
as schematically illustrated by the red arrows in Fig.~\ref{cryst}(a), since transferred hyperfine interactions are short-ranged.
This assumption is applicable for $^{1}$H nuclei since the distance from a H atom to the nearest Cu atom is 2.500 \AA, while that to the next-nearest Cu atom is 4.905 \AA.
For $^{23}$Na nuclei, the distance between Na and O is important since the transferred hyperfine interactions are mediated by Cu-O-Na paths.
The distance for the shortest path is 2.321 \AA \ and is considerably smaller than that for the next-shortest path of 2.806 \AA.
Thus, the assumption would be reasonable for $^{23}$Na nuclei as well.

The transferred hyperfine coupling used to analyze NMR spectra and $1/T_1$ is listed in Table~\ref{hyperfine} as $\sum_i \mathbf{A}^i_\mathrm{tr, sim}$.
While the transferred contribution for $^{23}$Na nuclei is almost isotropic, that for $^{1}$H nuclei is anistropic.
This anisotropy might be caused by the distribution of the magnetic moments over ligand O atoms due to the covalent bonding between Cu 3d and O 2s/2p orbitals,
which modifies the dipolar contribution.
Indeed, in several other compounds, the calculation of a hyperfine coupling constant is improved by putting a fraction of the magnetic moments on the ligand O atoms\cite{PbCuSO4OH_7, hyperfine, hyperfine2}.
However, in this compound, the remaining anistropy of hyperfine coupling cannot be reproduced by the same method.
Thus, we adopt the values determined under the assumption that the moments are only on Cu sites.

\begin{figure}[t]
\includegraphics[width=8.5cm]{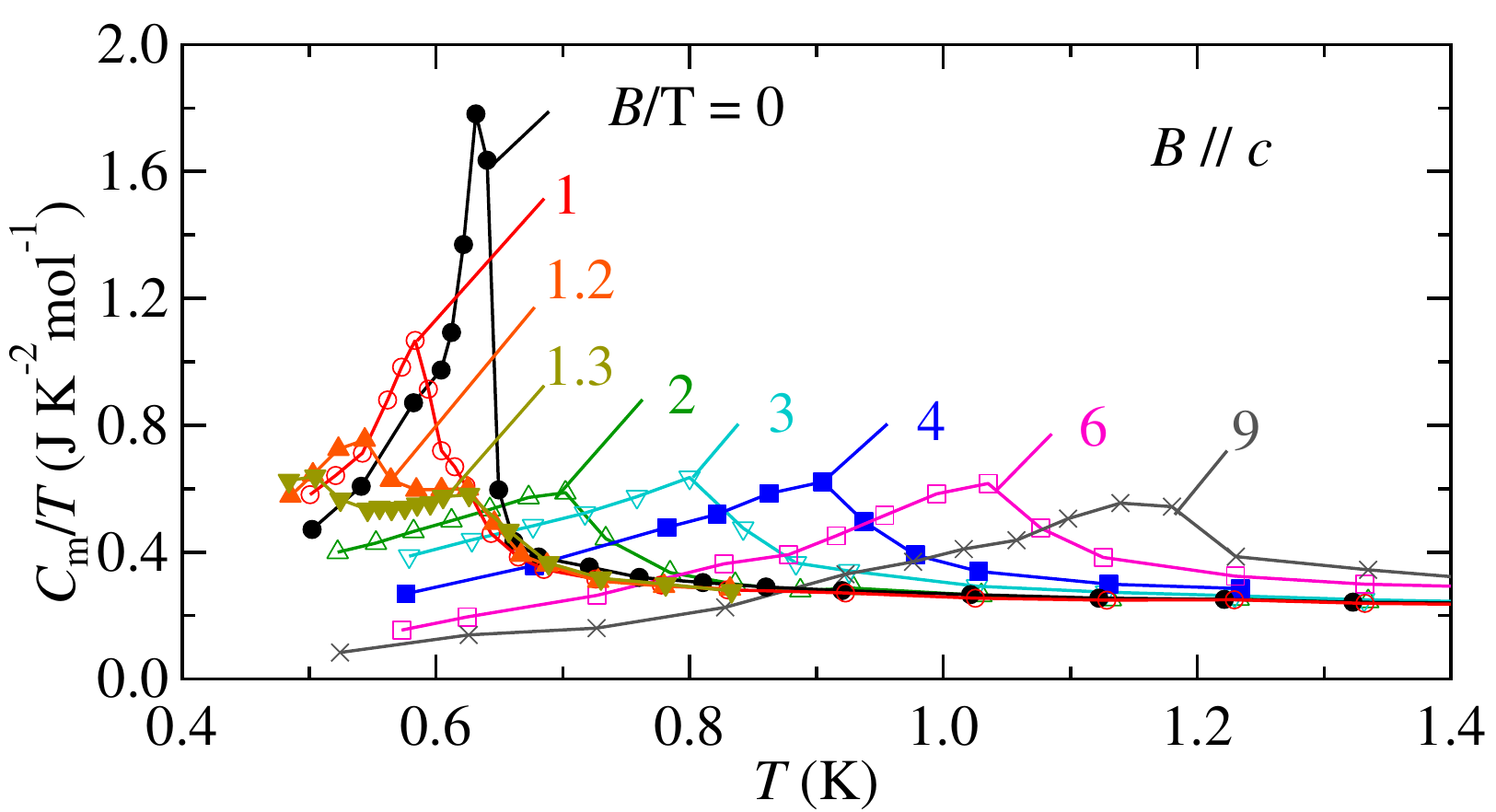}
\caption{\label{fig:TN} Temperature dependences of $C_m/T$ for $B~\parallel~c$.}
\end{figure}

\subsection{\label{PD}Phase diagram}
Before discussing magnetic structures and spin fluctuations, let us start with variations of the magnetic heat capacity $C_m$ and $^{23}$Na NMR spectra in order to establish a magnetic phase diagram.
The temperature dependence of $C_m/T$ at $B~\parallel~c$ is shown in Fig.~\ref{fig:TN}.
A sharp peak is observed at 0.63(1) K in zero field, indicating a magnetic phase transition\cite{NaCuMoO4OH}.
With an increasing magnetic field, the peak shifts to lower temperatures and splits into two peaks above 1~T.
The low-$T$ peak continues to move to lower temperatures and disappears below 0.5~K above 2~T, whereas the high-$T$ peak shifts to higher temperatures and finally reaches 1.16~K at 9~T.
These field dependences suggest the presence of two phases at low fields, which is confirmed by $^{23}$Na NMR measurements.

\begin{figure}[t]
\includegraphics[width=8.5cm]{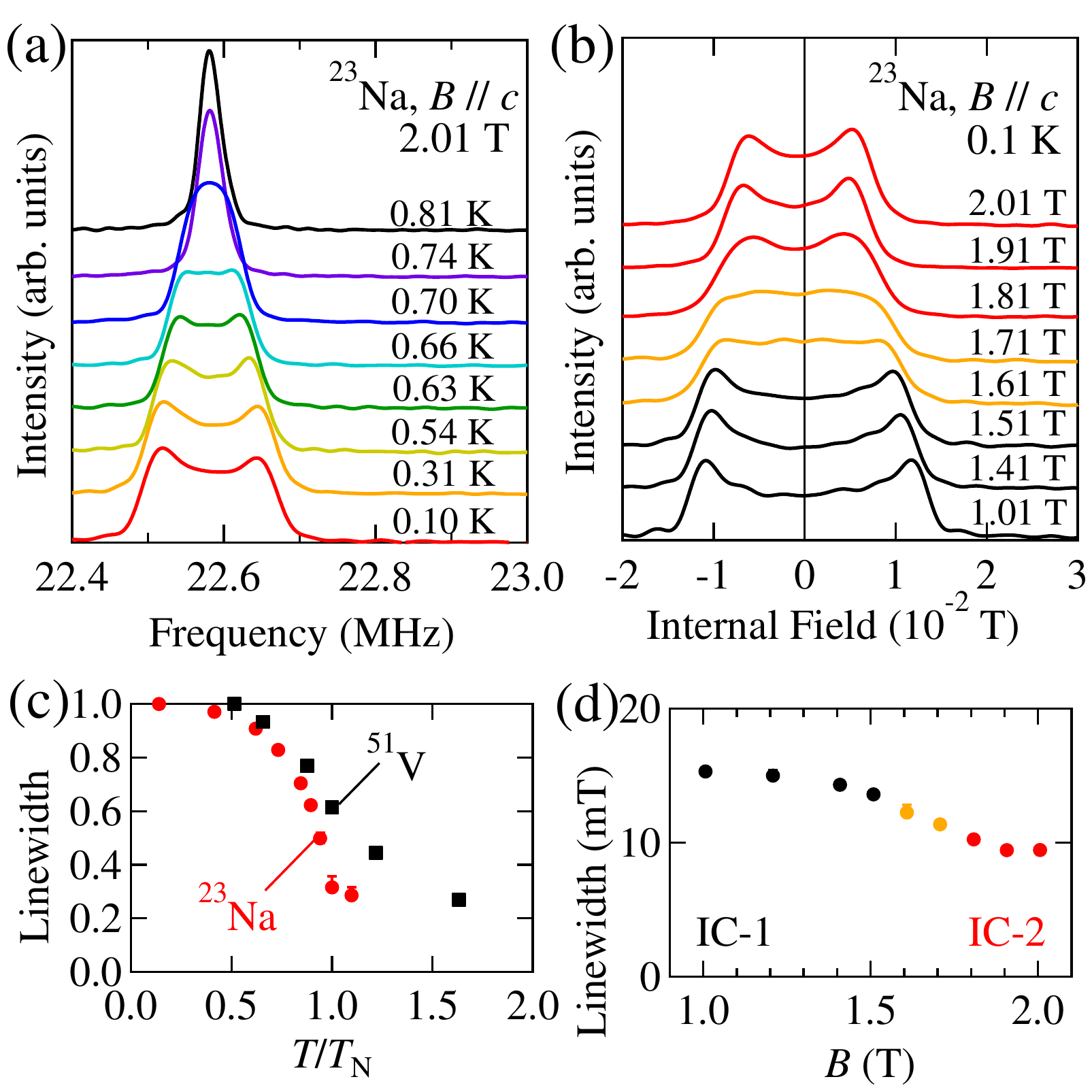}
\caption{\label{linewidth} (a) Temperature evolution of $^{23}$Na NMR spectra (center line) at $B$ =~2.01 T and $B~\parallel~c$.
(b) Variation of $^{23}$Na NMR spectra (center line) as a function of a magnetic field at $T$ = 0.1~K and $B~\parallel~c$.
(c) Normalized linewidth of the $^{23}$Na NMR spectra shown in Fig.~\ref{linewidth}(a) as a function of temperature, which is estimated from a second moment (red circles).
It is compared with the linewidth of $^{51}$V NMR spectra measured at $B$ = 10~T and $B~\parallel~c$ for $\mathrm{LiCuVO_4}$ (black squares) \cite{NMR6}.
(d) Field dependence of a linewidth estimated from $^{23}$Na NMR spectra shown in Fig.~\ref{linewidth}(b).
The linewidths in Fig.~(c, d) are determined by calculating second moments.
}
\end{figure}

Figure~\ref{linewidth}(a) shows the temperature dependence of $^{23}$Na NMR spectra at 2~T.
A sharp peak observed at 0.8~K and 2~T clearly becomes broad at lower temperatures.
The temperature dependence of the linewidth is shown in Fig.~\ref{linewidth}(c).
$T_\mathrm{N} =$ 0.7 K, determined by the onset temperature for line broadening, coincides with the peak temperature at 2~T in $C_m/T$.
The spectrum at 0.1~K shows a double-horn type lineshape, which is characteristic of an incommensurate spiral or SDW order.
Figure~\ref{linewidth}(b) shows a field evolution of NMR spectra.
A double-horn-type lineshape is also observed under lower magnetic fields.
Their linewidths are plotted as a function of a magnetic field in Fig~\ref{linewidth}(d).
A clear change in the linewidth is detected across $B_c$ =~1.51--1.81~T, indicating a field-induced magnetic phase transition between two incommensurate phases;
we name the two phases below and above $B_c$ as IC-1 and IC-2, respectively.
The transition between the two phases is observed at $B_c$ =~1.81--2.01~T for $B \parallel a$.
The difference in $B_c$ can be explained by the anisotropy of the $g$-factor\cite{NaCuMoO4OH3}.

\begin{figure}[t]
\includegraphics[width=8.5cm]{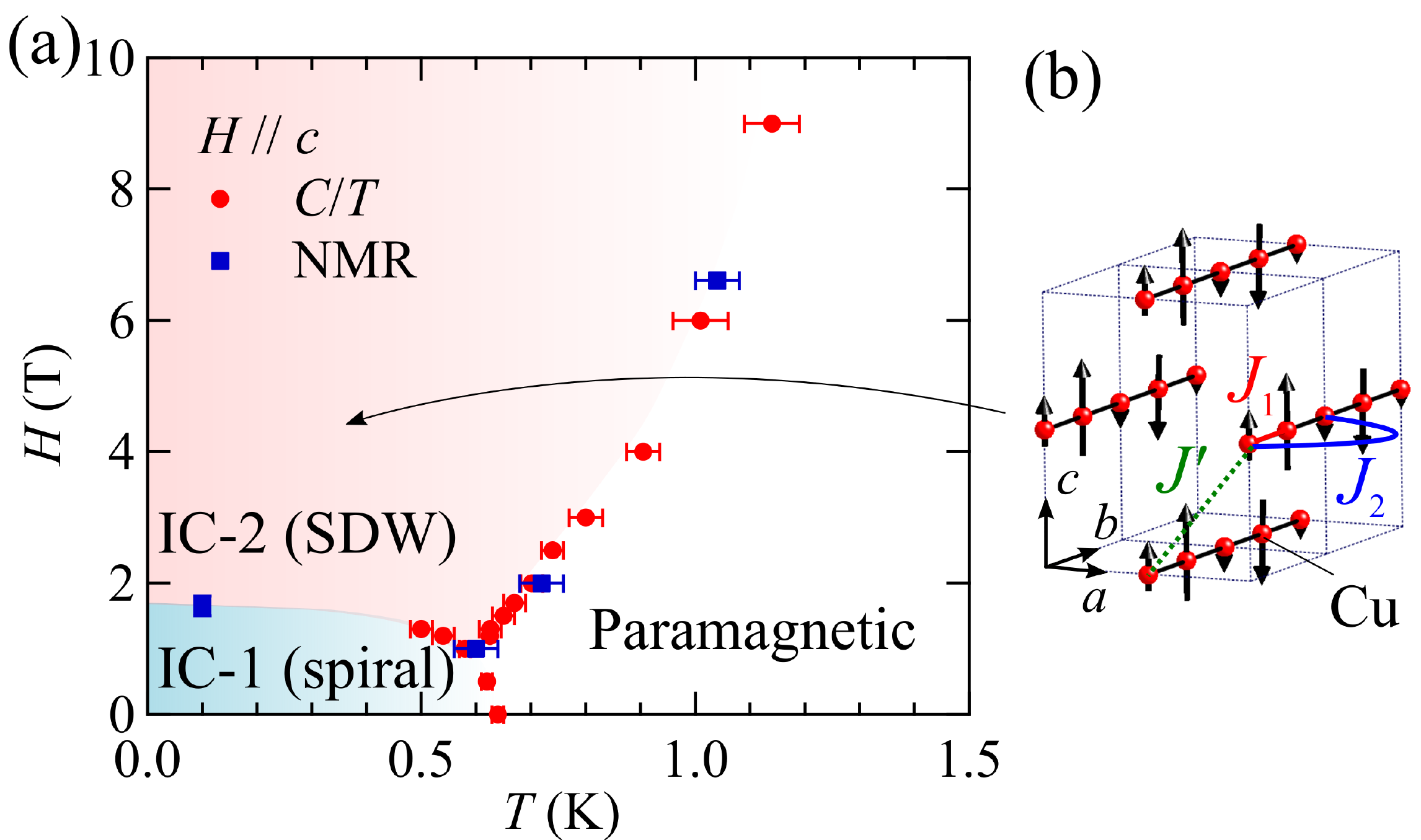}
\caption{\label{fig:phase_diagram} (a) Magnetic phase diagram of $\mathrm{NaCuMoO_4(OH)}$ for $B~\parallel~c$.
Two sets of $T_\mathrm{N}$ from heat capacity and $^{23}$Na NMR measurements are plotted.
(b) Schematic view of the magnetic structure in the IC-2 phase and dominant interactions.}
\end{figure}

All $T_\mathrm{N}$ from the heat-capacity and NMR measurements are plotted in the $B$-$T$ phase diagram of Fig.~\ref{fig:phase_diagram}(a).
IC-1 is quickly suppressed by $B$, while IC-2 becomes stable above $B_c$ with its $T_\mathrm{N}$ increasing with an increasing magnetic field.
Provided that the present compound is best described as a $J_1$--$J_2$ chain magnet, IC-1 and IC-2 would correspond to spiral and SDW phases,
respectively \cite{1Dtheory2, 1Dtheory3}.
DMRG calculations of a $J_1$--$J_2$ chain model show that the corresponding critical field is 0.05 $J_2$ for $J_1/J_2$ = $-$51/36 \cite{1Dtheory2, 1Dtheory3},
which corresponds to 1.2~T, reasonably close to the observed $B_c$.

\begin{figure*}[t]
\includegraphics[width=18cm]{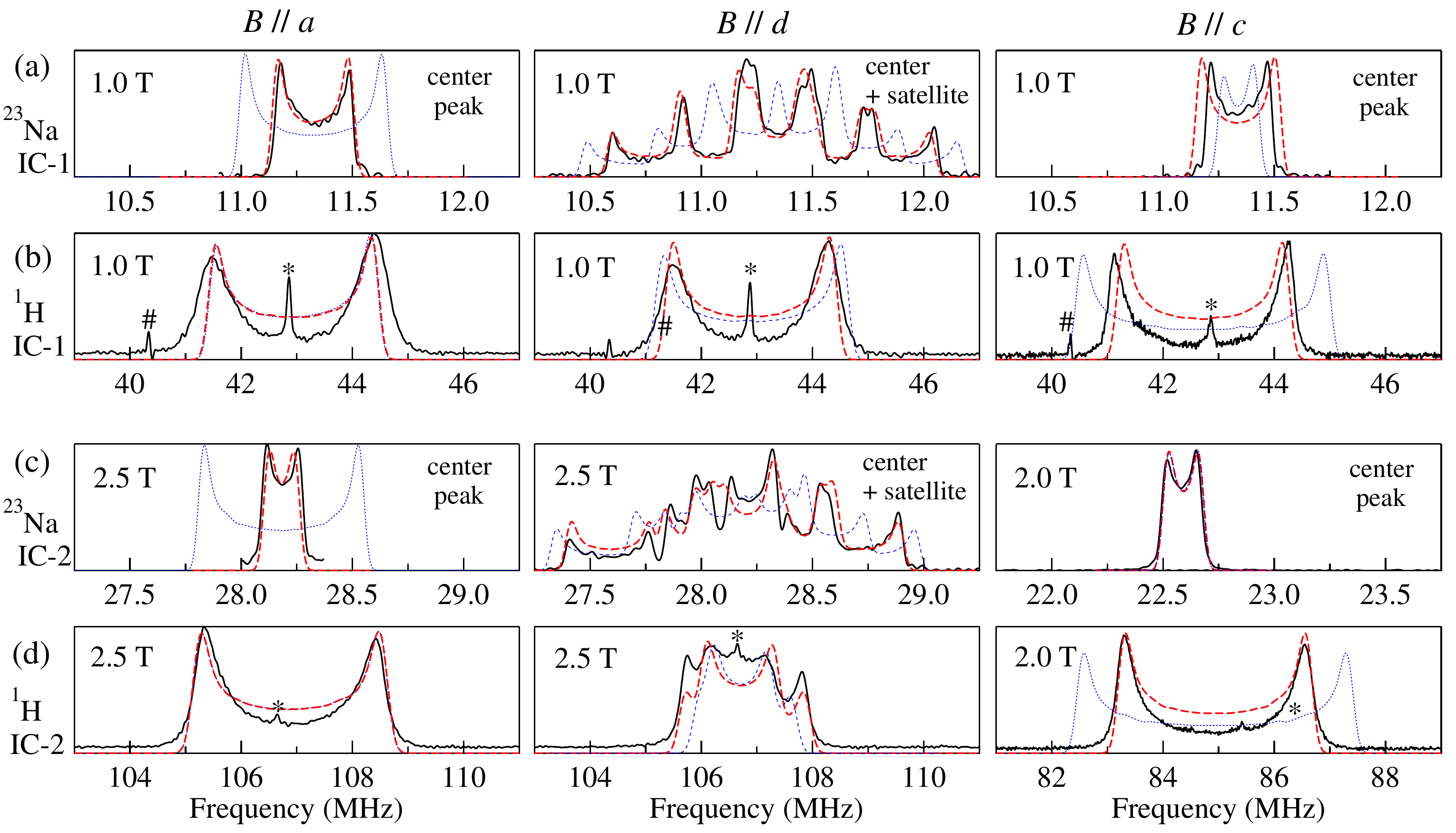}
\caption{\label{fig:spectra} $^{23}$Na and $^1$H NMR spectra for IC-1 (a, b) and IC-2 (c, d) measured at 0.1~K.
The left, middle, and right panels show the NMR spectra for $B~\parallel~a$, $B~\parallel~d$, and $B~\parallel~c$, respectively,
where the $d$ direction is canted from the $c$-axis to the $a$-axis by 61$^\circ$ (see Fig.~\ref{rotation}).
The left and right panels of (a) and (c) show only center peaks of the $^{23}$Na NMR signals since satellite peaks exhibit the same lineshape as the center one.
For the middle panels, two satellite peaks are also included since they cannot be separated from the center line.
An analytic deconvolution is applied to the quadrupole splitting of the NMR spectra in the top panel of (a) \cite{deconvolution}.
In each panel, the experimental NMR spectrum (black solid curves) is compared with simulated spectra for the spiral (IC-1) or SDW (IC-2) order 
with the ferromagnetic (red dashed curves) and antiferromagnetic (blue dotted curves) interchain coupling $J'$.
Symbols * and \# represent extrinsic signals from $^1$H and $^{19}$F nuclei, respectively.}

\end{figure*}

\subsection{\label{order}Magnetic structures at ordered phases} 
To determine the magnetic structures of IC-1 and IC-2, we carefully performed $^{23}$Na and $^1$H NMR measurements with the three orientations of $B~\parallel~a$, $B~\parallel~d$, and $B~\parallel~c$,
where the $d$ direction is canted from the $c$ axis to $a$ axis by 61$^\circ$.
The $d$ direction is selected so that one set of a magnetic shift for $^1$H nuclei becomes almost 0, as shown in Fig.~\ref{rotation}(b).
The obtained spectra are shown by the black solid curves in Fig.~\ref{fig:spectra}.
For $B~\parallel~a$ and $c$ (the left and right panels), there is a unique site either for a Na or H atom in the paramagnetic state
so that an incommensurate magnetic order produces a single resonance line with a double-horn structure.
On the other hand, the NMR spectra for $B~\parallel~d$ (the middle panels) can be complex
because two inequivalent sites are present either for a Na or H atom, unless for $B \parallel ab$ or $bc$, which lead to the overlap of the two double-horn lineshapes.
Such a complex $B~\parallel~d$ spectrum could be decisive in determining the spin structure.
Note that a $^{23}$Na NMR spectrum also contains two satellite peaks together with the center peak, thus totally a superposition of six double-horn lineshapes appears.

First, we discuss the $^1$H NMR spectra in IC-1 (Fig.~\ref{fig:spectra}(b)) and IC-2 (Fig.~\ref{fig:spectra}(d)). 
While the $^1$H NMR spectra in IC-1 are insensitive to the applied field direction, the spectral width in IC-2 is strongly dependent on the field direction.
This field-direction dependence in IC-2 agrees well with the angular dependence of the paramagnetic shift shown in Fig.~\ref{rotation}(b),
indicating that the ordered moments in IC-2 are parallel to the field direction.
Thus, the magnetic structure in IC-2 is considered to be SDW, as expected in the $J_1$--$J_2$ chain.
On the other hand, it is difficult to deduce the magnetic structure for IC-1, where a spiral order  is expected.
This is because the transverse ordered moments combined with the off-diagonal component of the hyperfine coupling can also contribute to the internal field,
and thus, the angular dependence of the NMR spectra for a spiral order is not straightforward.

To examine details of the magnetic structures,
we performed a simulation of the spectra by constructing a histogram of the resonance frequency $\nu$ = $\gamma |\mathbf{B} + \mathbf{B}_\mathrm{int}|$
and then convoluting it with a Gaussian function.
To obtain the distribution of $\nu$, the internal field $\mathbf{B}_\mathrm{int}$ is calculated as $\mathbf{B}_\mathrm{int}$ = $\sum_i \mathbf{A}^i \cdot \mathbf{\mu}^i$,
where $\mathbf{A}^i$ is the hyperfine coupling tensor discussed in Section~\ref{para} and $\mathbf{\mu}^i$ is the magnetic moment of an assumed spin structure at the i-th Cu site within a distance of 60~\AA \ from the nuclei.
Note that the $ab$- and $bc$-components of the transferred hyperfine coupling, which cannot be determined in the paramagnetic phase, are set to zero.
These components have almost no influence on our final result, since there is no ordered moment along the $b$-axis.
The NMR spectra could not be reproduced by spiral structures in the $ab$- or $bc$-plane even if $A_{ab}$ and $A_{bc}$ are treated as adjustable parameters.

For IC-2, the magnetic structure is expected to be an SDW order structure with spins aligned parallel to the magnetic field and modulated sinusoidally along the spin chain.
We have performed simulations for two cases: the case of ferromagnetic interchain coupling (defined as $J'$ in Fig.~\ref{fig:phase_diagram}(b)) and of antiferromagnetic interchain coupling.
The magnetic wave vectors of the two cases are $\mathbf{Q} = 2 \pi (0, \alpha, 0)$ and $2 \pi (1, \alpha, 0)$, respectively,
where $\alpha$ is $\alpha = 1/2 -  M/(g \mu_B)$ ($M$ is the magnetization) deduced from the $J_1$--$J_2$ chain model \cite{1Dtheory2, 1Dtheory3, 1DtheoryofT11};
note that the unit cell includes two Cu sites in a single chain.
As shown in Figs.~\ref{fig:spectra}(c, d), the simulation for ferromagnetic $J'$ (red dashed curves) can reproduce all of the experimental spectra,
whereas that for antiferromagnetic $J'$ (blue dotted curves) cannot.
Thus, the SDW with ferromagnetic $J'$, as shown in Fig.~\ref{fig:phase_diagram}(b), is the most likely candidate.
Note that the amplitude of the SDW is the only free parameter except for $\mathbf{A}_\mathrm{tr}$, which has little uncertainty.
The amplitude is estimated to be 0.38~$\mu_\mathrm{B}$, assuming that it is independent of the field orientation.
It is smaller than the value of 0.6--0.8~$\mu_\mathrm{B}$ for $\mathrm{LiCuVO_4}$ \cite{NMR4},
which may be due to larger quantum fluctuations associated with better one dimensionality in $\mathrm{NaCuMoO_4(OH)}$.

On the other hand, we have examined four likely cases for IC-1: the spiral plane always perpendicular to the field direction or parallel to the $ab$-, $bc$-, or $ac$-plane regardless of the field direction.
The magnetic wave vector is $\mathbf{Q} = 2 \pi (0, \beta, 0)$, where $\beta$ is $\beta = \arccos(-J_1/4 J_2)/\pi$ in a classical $J_1$--$J_2$ chain.
Among the four cases, all of the experimental spectra are well reproduced only when the spiral plane is parallel to the $ac$-plane,
as shown in Figs.~\ref{fig:spectra}(a, b).
Only the $ac$-spiral order with ferromagnetic $J'$ is consistent with the experimental spectra.
The magnitude of the ordered moments is estimated to be 0.29 $\mu_\mathrm{B}$.
Note that this value, based on the classical model, may be an underestimation since quantum effects should lead to a larger pitch angle of the spiral,
resulting in narrower NMR spectra.
In brief summary, the NMR spectra indicate the spiral and SDW orders in IC-1 and IC-2, respectively,
as expected from the frustrated $J_1$--$J_2$ chain model.

\subsection{\label{T1subsec}Anisotropic spin fluctuations} 
\begin{figure}[t]
\includegraphics[width=8.5cm]{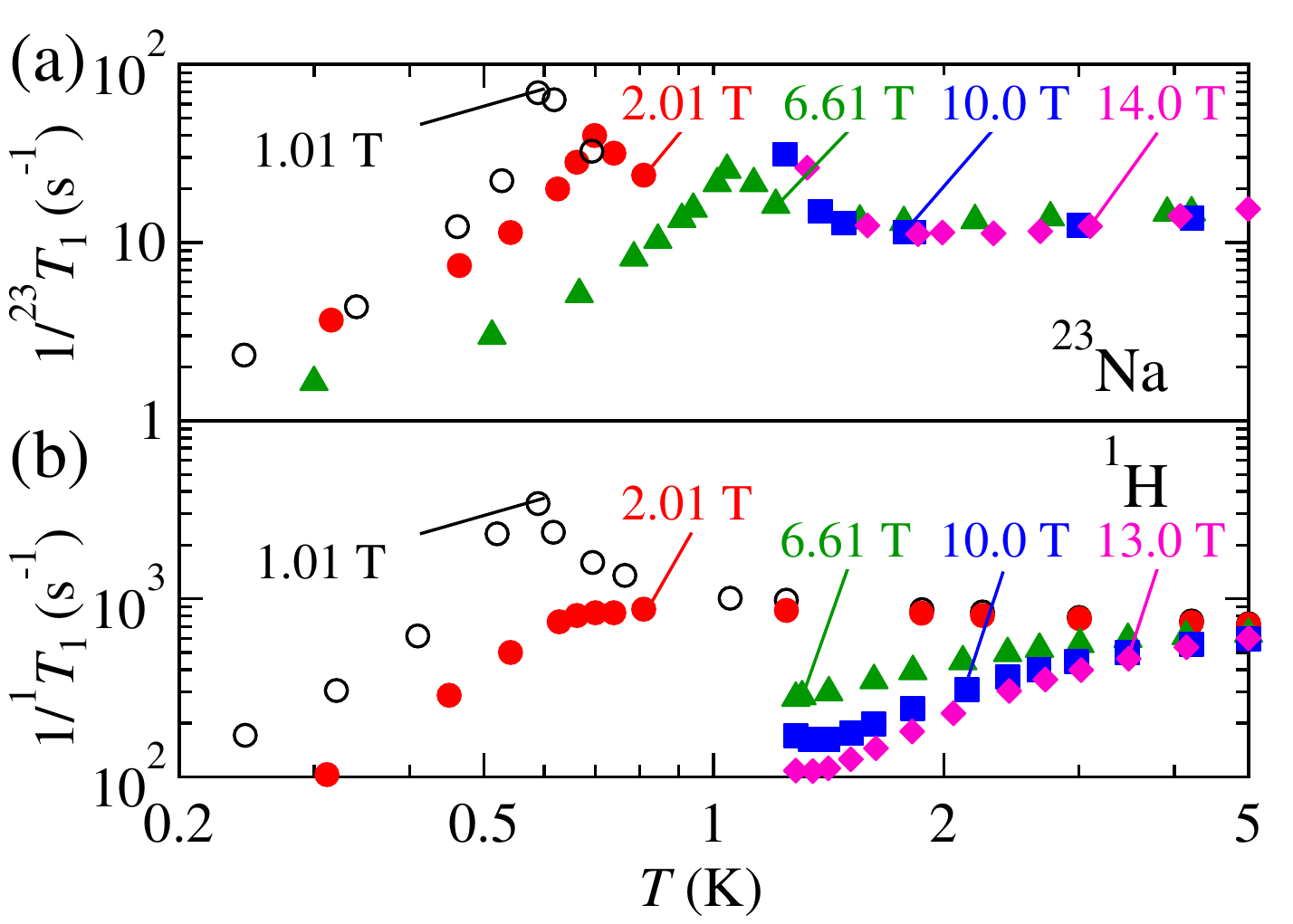}
\caption{\label{fig:T1} Temperature dependences of $1/T_1$ at $B \parallel c$ for (a) $^{23}$Na and (b) $^1$H nuclei.}
\end{figure}

\begin{figure}[t]
\includegraphics[width=8.5cm]{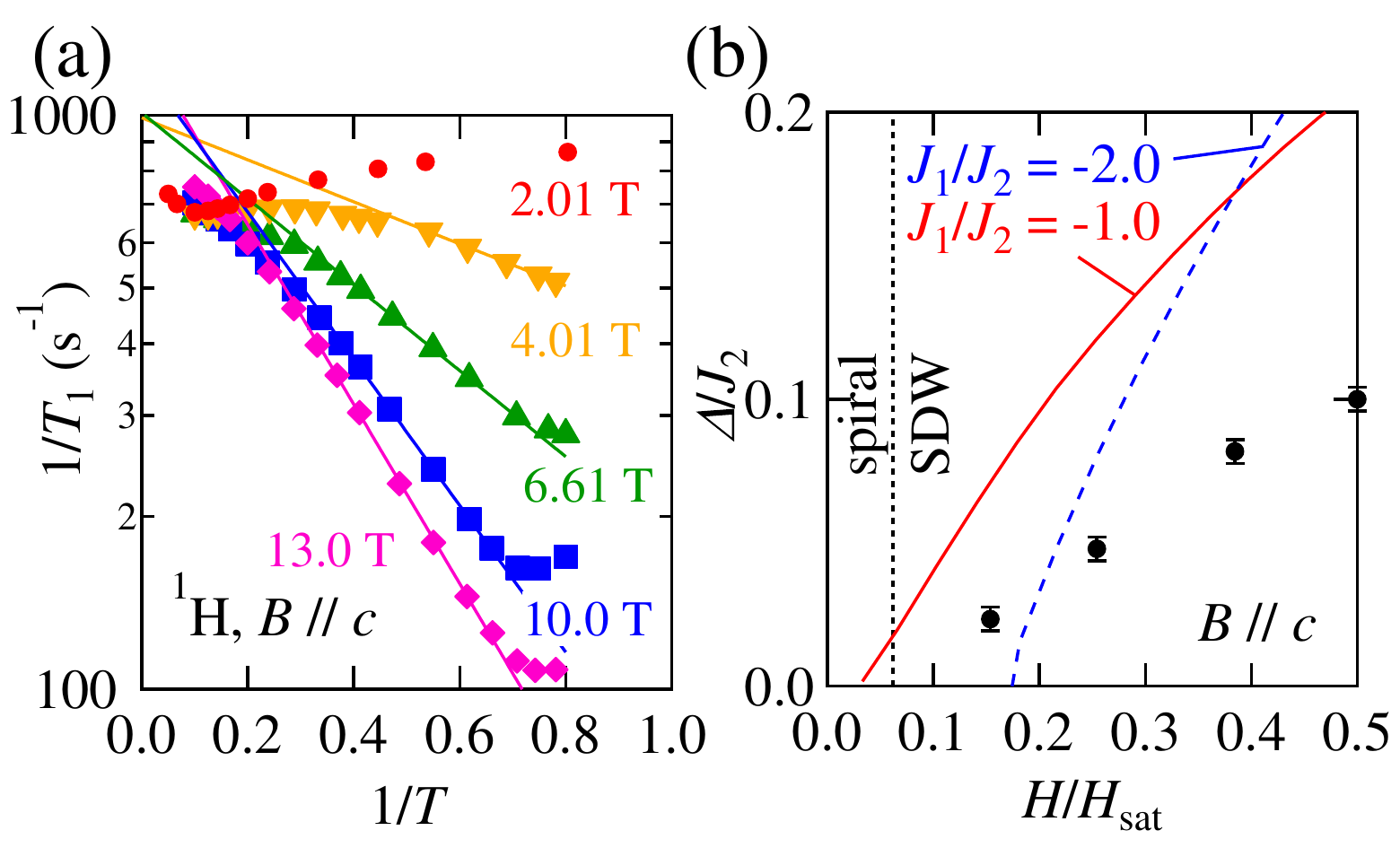}
\caption{\label{fig:gap} (a) Arrhenius plot of $1/^1T_1$ measured at 2--13~T.
(b) Field dependence of the excitation gap determined from $1/^1T_1$.
Red solid and blue dashed curves represent field dependences of magnon binding energy in a frustrated $J_1$-$J_2$ chain with $J_2/J_1$ = $-$1.0 and $-$2.0 calculated from DMRG calculations\cite{1Dtheory5}.}
\end{figure}

Another evidence for the $J_1$--$J_2$ chain magnet is found in the presence of anisotropic spin fluctuations due to
the formation of magnon bound states.
Figures~\ref{fig:T1}(a) and \ref{fig:T1}(b) show the temperature dependences of $1/T_1$ at $B \parallel c$ for $^{23}$Na ($1/^{23}T_1$) and $^{1}$H ($1/^1T_1$), respectively.
$1/^{23}T_1$ behaves similarly below and above $B_c$;
it increases with decreasing temperature and exhibits a peak at $T_\mathrm{N}$ owing to the critical slowing down of spin fluctuations.
In sharp contrast, $1/^1T_1$ changes its temperature dependence remarkably across $B_c$:
the enhancement in $1/^1T_1$ observed at 1.01~T near $T_\mathrm{N}$  is suppressed at 2.01~T just above $B_c$.
At higher magnetic fields, $1/^1T_1$ decreases with decreasing temperature and follows an activation-type temperature dependence above $T_\mathrm{N}$.
This is confirmed by an Arrhenius plot of $1/^1T_1$ in Fig.~\ref{fig:gap}(a). 
At 10 T, the activation energy is estimated to be $\Delta$ = 2.9(1) K $\sim$ 0.08 $J_2$.

In order to understand the difference between the temperature dependences of $1/^{23}T_1$ and $1/^1T_1$, it is necessary to investigate the form factor for both nuclei.
In general, $(1/T_1)_\xi$, where $\xi$ denotes the field direction, is given by
the sum of both transverse and longitudinal spin correlation functions $S_\perp (\mathbf{q}, \omega)$ and $S_\parallel (\mathbf{q}, \omega)$\cite{NMR2}:
\begin{equation}
\begin{split}
\left( \frac{1}{T_1} \right)_\xi &= \frac{1}{N} \sum_\mathbf{q} \{\Gamma^{\perp}_\xi(\mathbf{q}) S_\perp (\mathbf{q}, \omega)
+ \Gamma^{\parallel}_\xi(\mathbf{q}) S_\parallel (\mathbf{q}, \omega) \},
\label{T1}
\end{split}
\end{equation}
where $N$ is the number of atoms, and $\Gamma^{\perp}_\xi(\mathbf{q})$ and $\Gamma^{\parallel}_\xi(\mathbf{q})$
are form factors defined as in Ref.~\onlinecite{NMR2}.
For $B \parallel c$, they become
\begin{equation}
\begin{split}
\Gamma^{\perp}_c(\mathbf{q}) &= \frac{\gamma^2}{2} \{ g_{aa}^2 | A(\mathbf{q})_{aa} |^2 + g_{bb}^2 | A(\mathbf{q})_{bb} |^2 \\
&\ \ \ \ + (g_{aa}^2 + g_{bb}^2) | A(\mathbf{q})_{ab} |^2 \} \\
\Gamma^{\parallel}_c(\mathbf{q}) &= \frac{\gamma^2}{2} g_{cc}^2 \left(  | A(\mathbf{q})_{ac} |^2 + | A(\mathbf{q})_{bc} |^2 \right),
\label{form factor}
\end{split}
\end{equation}
where $A(\mathbf{q})_{\mu\nu}$ is a Fourier sum of hyperfine coupling constants, $A(\mathbf{q})_{\mu\nu} = \sum_i A^i_{\mu\nu} e^{i\mathbf{q} \cdot \mathbf{r}}$,
taken over all Cu sites within a distance of 60 \AA \ from the nuclei.
In a small temperature range just above $T_\mathrm{N}$,
where spin fluctuations are dominated by the component with the $\mathbf{q}$-vector in the ordered phase $\mathbf{Q}_0$, 
the $q$-dependent hyperfine coupling constants in Eq.~\eqref{T1} can be approximately replaced by their values at $\mathbf{Q}_0$\cite{NMR2}:
\begin{equation}
\left( \frac{1}{T_1} \right)_\xi \simeq \Gamma^{\perp}_\xi(\mathbf{Q}_0) \langle S_\perp (\mathbf{q}, \omega) \rangle 
+ \Gamma^{\parallel}_\xi(\mathbf{Q}_0) \langle S_\parallel (\mathbf{q}, \omega) \rangle,
\label{T1uni}
\end{equation}
where $\langle S_\perp (\mathbf{q}, \omega) \rangle$ and $\langle S_\parallel (\mathbf{q}, \omega) \rangle$ represent q-averages of
the transverse and longitudinal spin correlation functions, respectively.

Equation~\eqref{T1uni} indicates that $\langle S_\perp (\mathbf{q}, \omega) \rangle$ and $\langle S_\parallel (\mathbf{q}, \omega) \rangle$ close to $T_N$
can be extracted by calculating $\Gamma^{\perp}_c \equiv \Gamma^{\perp}_\xi(\mathbf{Q}_0)$ and $\Gamma^{\parallel}_c \equiv  \Gamma^{\parallel}_\xi(\mathbf{Q}_0)$
from the hyperfine coupling tensor and the magnetic wave vector $\mathbf{Q}_0$ = $2 \pi (0, \alpha, 0)$.
We adopt the transferred hyperfine coupling constants listed in Table~\ref{hyperfine} for this calculation.
The $ab$- and $bc$-components of the transferred hyperfine coupling tensor, which cannot be determined experimentally, are assumed to be zero.
For $^{23}$Na nuclei, $\Gamma^{\perp}_c$ and $\Gamma^{\parallel}_c$ are estimated as
7.5 $\times$ 10$^{13}$ and 3.8 $\times$ 10$^{13}$ s$^{-2}$ at 2 T, respectively, leading to $\Gamma^{\perp}_c/\Gamma^{\parallel}_c$ = 2.0.
Thus, both the transverse and longitudinal spin fluctuations affect $1/^{23}T_1$.
On the other hand, the same procedure provides a $\Gamma^{\perp}_c$ much larger than $\Gamma^{\parallel}_c$ for $^1$H nuclei:
6.2 $\times$ 10$^{15}$ and 1.0 $\times$ 10$^{14}$ s$^{-2}$, respectively ($\Gamma^{\perp}_c/\Gamma^{\parallel}_c$ = 60).
This is because H and Cu atoms are almost in the same $c$-plane, and thus, dominant dipole-dipole interactions provide $|A_{ac}|$ and $|A_{bc}|$ much smaller than $|A_{aa}|$ and $|A_{ab}|$.
The large $\Gamma^{\perp}_c/\Gamma^{\parallel}_c$ indicates that $1/^1T_1$ is only sensitive to transverse fluctuations.
Based on both form factors, we come to the following conclusion: the activated temperature dependence in $1/^1T_1$ reveals the presence of gapped transverse excitations,
while the strong increase near $T_\mathrm{N}$ in $1/^{23}T_1$ indicates gapless longitudinal excitations.
In addition, the above conclusion is not changed by the uncertainty of $A_{ab}$ and $A_{bc}$. 
Even if an additional contribution of $A_\mathrm{tr}$ comparable with $A_\mathrm{dip}$ is added in the Fourier sum,
$\Gamma_c^\perp/\Gamma_c^\parallel \sim 1$ for $^{23}$Na and $\Gamma_c^\perp/\Gamma_c^\parallel \gg 1$ for $^1$H are still satisfied.

Such anisotropic spin fluctuations are consistent with the formation of bound magnons expected in the $J_1$--$J_2$ chain magnet.
The gap cannot be explained by the Zeeman energy, since it induces a gap in a longitudinal spectrum, which is inconsistent with the anisotropic gap in this compound.
The gap corresponds to the magnon binding energy, which is the energy cost to separate a magnon bound pair into two single magnons, resulting in gapped transverse excitations \cite{1DtheoryofT11, 1DtheoryofT12}.
At the same time, longitudinal fluctuations are developed because of density fluctuations of bound magnons.
The field dependence of the gap estimated from the Arrhenius plot is compared with the magnon binding energy in a frustrated $J_1$-$J_2$ chain determined from DMRG calculations in Fig.~\ref{fig:gap}(b)\cite{1Dtheory5}.
The gap becomes large with increasing field, which is qualitatively consistent with the field dependence of the magnon binding energy.
However, its magnitude is almost half of that of the $J_1$-$J_2$ chain model. 
This may be due to Dzyaloshinskii-Moriya interactions not included in the DMRG calculation,
which can be the same magnitude as the magnon binding energy.
Note that Dzyaloshinskii-Moriya interactions between nearest neighbors are present while those between next-nearest neighbors are absent because of inversion symmetry at each Cu site.

\section{\label{dis}Comparison with other candidates}
The present study reveals that $\mathrm{NaCuMoO_4(OH)}$ realizes a $J_1$--$J_2$ chain magnet from both macroscopic and microscopic probes.
Compared with other candidates, the magnetic properties of $\mathrm{NaCuMoO_4(OH)}$ are quite similar to those of $\mathrm{LiCuVO_4}$.
For instance, the phase diagram of these materials has the same character at low fields:
the spiral and collinear SDW phases are present, and the transition temperature of the spiral phase decreases but that of the collinear SDW phase increases with an increasing field\cite{PD_LCVO, NMR4, HFHC}.
Additional intermediate phases triggered by competition among interchain interactions,
such as a complex collinear SDW phase in $\mathrm{PbCu(SO_4)(OH)_2}$\cite{PbCuSO4OH_3, PbCuSO4OH_6, PbCuSO4OH_8} and a spin-stripe phase in $\mathrm{TeVO_4}$\cite{TeVO4_3, TeVO4_4}, have not been detected so far. 
The difference indicates that interchain interactions are weak in $\mathrm{NaCuMoO_4(OH)}$.

The difference between $\mathrm{NaCuMoO_4(OH)}$ and $\mathrm{LiCuVO_4}$ is that the former compound has a great advantage to obtain high-quality single crystals with less disorder 
as well as $\mathrm{PbCu(SO_4)(OH)_2}$\cite{PbCuSO4OH_0, PbCuSO4OH_2, PbCuSO4OH_3, PbCuSO4OH_4, PbCuSO4OH_6, PbCuSO4OH_7, PbCuSO4OH_8, PbCuSO4OH_9},
3-I-V\cite{3IV}, and $\mathrm{TeVO_4}$\cite{TeVO4, TeVO4_2, TeVO4_3, TeVO4_4},
while the latter compound has difficulties in avoiding disorder effects.
$C_m/T$ and the linewidth of NMR spectra in the vicinity of $T_\mathrm{N}$ can be indicators of the degree of disorder.
$C_m/T$ exhibits a sharp peak at $T_\mathrm{N}$ in $\mathrm{NaCuMoO_4(OH)}$ (Fig.~\ref{fig:TN}),
in contrast to the much broader peak in $\mathrm{LiCuVO_4}$\cite{neutron2, HFHC}.
Moreover, the second moment of NMR spectra exhibits an abrupt change in $\mathrm{NaCuMoO_4(OH)}$ below $T_\mathrm{N}$,
while it exhibits a broad variation in the $^{51}$V NMR spectra of $\mathrm{LiCuVO_4}$ \cite{NMR6}, as compared in Fig.~\ref{linewidth}(c).
The availability of high-quality crystals is important since the nematic state might be significantly suppressed by disorder, especially in high fields,
and the transition should be sharp to detect the nematic state expected in the very narrow field range.
In addition, the smaller saturation field of $\mathrm{NaCuMoO_4(OH)}$ compared to that of $\mathrm{LiCuVO_4}$ makes high-field experiments easier.
From these viewpoints, $\mathrm{NaCuMoO_4(OH)}$ is a promising compound to investigate unique field-induced phases in the $J_1$--$J_2$ chain magnet.

\section{\label{summary}Summary} 
In summary, we performed heat-capacity and NMR measurements on a single crystal of $\mathrm{NaCuMoO_4(OH)}$.
A magnetic-field-induced transition is found at $B_c$ $\sim$~1.8~T from an incommensurate spiral order to an incommensurate longitudinal SDW order
in which anisotropic spin fluctuations indicating the formation of bound magnons are observed by $1/T_1$ measurements.
Therefore, $\mathrm{NaCuMoO_4(OH)}$ is a good candidate frustrated $J_1$--$J_2$ chain magnet and
would provide us an opportunity to investigate the hidden spin nematic order and fluctuations near the magnetic saturation
through further high-field NMR and neutron scattering experiments.

\begin{acknowledgments}
We thank O. Janson for DFT calculations and N. Shannon, T. Masuda, S. Asai, and T. Oyama for fruitful discussions.
This work was supported by a Grant-in-Aid for Young Scientists (B) (No. 15K17693, No. 26800176).
\end{acknowledgments}


\begin{thebibliography}{99}
\bibitem{SL} H.-J. Mikeska and A. K. Kolezhuk, in \textit{Quantum Magnetism}, edited by U. Schollw\"ock et al., Lecture Notes in Physics Vol. 645
(Springer-Verlag, Berlin, 2004), p. 1.
\bibitem{SL2} G. Misguich and C. Lhuillier, in \textit{Frustrated Spin Systems}, edited by H. T. Diep (World Scientific, Singapore, 2005), p. 229.
\bibitem{SL3} L. Balents, Nature \textbf{464}, 199 (2010).
\bibitem{VBC} M. E. Zhitomirsky and K. Ueda, Phys. Rev. B \textbf{54}, 9007 (1996).
\bibitem{nematic} F. Andreev and I. A. Grishchuk, Zh. Eksp. Teor. Fiz. \textbf{87}, 467 (1984).
\bibitem{nematic2} N. Shannon, T. Momoi, and P. Sindzingre, Phys. Rev. Lett. \textbf{96}, 027213 (2006).
\bibitem{nematic3} H. T. Ueda and T. Momoi, Phys. Rev. B \textbf{87}, 144417 (2013).
\bibitem{1Dtheory00} A. V. Chubukov, Phys. Rev. B \textbf{44}, 4693 (1991).
\bibitem{1Dtheory0} L. Kecke, T. Momoi, and A. Furusaki, Phys. Rev. B \textbf{76} 060407 (2007).
\bibitem{1Dtheory1} T. Vekua, A. Honecker, H.-J. Mikeska, and F. Heidrich-Meisner, Phys. Rev. B \textbf{76}, 174420 (2007).
\bibitem{1Dtheory2} T. Hikihara, L Kecke, T. Momoi, and A. Furusaki, Phys. Rev. B \textbf{78}, 144404 (2008).
\bibitem{1Dtheory3} J. Sudan, A. L\^uscher, and A. M. L\^auchli, Phys. Rev. B \textbf{80}, 140402 (2009).
\bibitem{1Dtheory4} M. E. Zhitomirsky and H. Tsunetsugu, Europhys. Lett. \textbf{92}, 37001 (2010).
\bibitem{1Dtheory5} M. Sato, T. Hikihara, and T. Momoi, Phys. Rev. Lett. \textbf{110}, 077206 (2013).
\bibitem{1Dtheory6} O. A. Starykh and L. Balents, Phys. Rev. B \textbf{89}, 104407 (2014).
\bibitem{1Dtheory7} L. Balents and O. A. Starykh, Phys. Rev. Lett. \textbf{116}, 177201 (2016).
\bibitem{LiCuVO4} M. A. Lafontaine, M. Leblanc, and G. Ferey, Inorg. Chem. \textbf{C45}, 1205 (1989).
\bibitem{neutron00} B. J. Gibson, R. K. Kremer, A. V. Prokofiev, W. Assmus, and G. J. McIntyre, Physica B \textbf{350}, e253 (2004).
\bibitem{neutron0} M. Enderle, C. Mukherjee, B. F\aa k, R. K. Kremer, J.-M. Broto, H. Rosner, S.-L. Drechsler, J. Richter, J. Malek, A. Prokofiev,
W. Assmus, S. Pujol, J.-L. Raggazzoni, H. Rakoto, M. Rheinst\^adter and H. M. R\o nnow, Europhys. Lett. \textbf{70}, 237 (2005).
\bibitem{NMR1} N. B\"{u}ttgen, H. -A. Krug von Nidda, L. E. Svistov, L. A. Prozorova, A. Prokofiev, and  W. A\ss mus,
Phys. Rev. B \textbf{76}, 014440 (2007).
\bibitem{PD_LCVO}
M. G. Banks, F. Heidrich-Meisner, A. Honecker, H. Rakoto, J. -M. Broto, and R. K. Kremer, J. Phys.: Condens. Matter, \textbf{19} 145227 (2007).
\bibitem{neutron1} T. Masuda, M. Hagihala, Y. Kondoh, K. Kaneko, and N. Metoki, 
J. Phys. Soc. Jpn. \textbf{80}, 113705 (2011).
\bibitem{neutron2} M. Mourigal, M. Enderle, B. F\aa k, R. K. Kremer, J. M. Law, A. Schneidewind, A. Hiess, and A. Prokofiev,
Phys. Rev. Lett. \textbf{109}, 027203 (2012).
\bibitem{NMR3} N. B\"{u}ttgen, W. Kraetschmer, L. E. Svistov, L. A. Prozorova, and A. Prokofiev,
Phys. Rev. B \textbf{81}, 052403 (2010).
\bibitem{NMR4} N. B\"{u}ttgen, P. Kuhns, A. Prokofiev, A. P. Reyes, and L. E. Svistov,
Phys. Rev. B \textbf{85}, 214421 (2012).
\bibitem{NMR2} K. Nawa, M. Takigawa, M. Yoshida, and K. Yoshimura, J. Phys. Soc. Jpn. \textbf{82}, 094709 (2013).
\bibitem{NMR5} K. Nawa, M. Takigawa, S. Kr\"{a}mer, M. Horvatic\', C. Berthier, M. Yoshida, and K. Yoshimura, Phys. Rev. B \textbf{96}, 134423 (2017).
\bibitem{magnetization} L. E. Svistov, T. Fujita, H. Yamaguchi, S. Kimura, K. Omura, A. Prokofiev, A. I. Smirnov, Z. Honda, and M. Hagiwara, 
JETP Lett. \textbf{93}, 21 (2011).
\bibitem{HFNMR}
N. B\"{u}ttgen, K. Nawa, T. Fujita, M. Hagiwara, P. Kuhns, A. Prokofiev, A. P. Reyes, L. E. Svistov, K. Yoshimura, and M. Takigawa, Phys. Rev. B \textbf{90}, 134401 (2014).
\bibitem{HFHC}
L. A. Prozorova, S. S. Sosin, L. E. Svistov, N. B\"{u}ttgen, J. B. Kemper, A. P. Reyes, S. Riggs, A. Prokofiev, and O. A. Petrenko,
Phys. Rev. B \textbf{91}, 174410 (2015).
\bibitem{HFNMR2}
A. Orlova, E. L. Green, J. M. Law, D. I. Gorbunov, G. Chanda, S. Kr\"amer, M. Horvati\'c, R. K. Kremer, J. Wosnitza, and G. L. J. A. Rikken,
Phys. Rev. Lett. \textbf{118}, 247201 (2017).
\bibitem{Li2ZrCuO4_0} C. Dussarrat, G. C. Mather, V. Caignaert, B. Domen\`es, J. G. Fletcher, and A. R. West, J. Solid. State Chem. \textbf{166}, 311 (2002).
\bibitem{Li2ZrCuO4} S.-L. Drechsler, O. Volkova, A. N. Vasiliev, N. Tristan, J. Richter, M. Schmitt, H. Rosner, J. M\'alek, R. Klingeler, A. A. Zvyagin,
and B. Bu\"chner, Phys. Rev. Lett. \textbf{98}, 077202 (2007).
\bibitem{Rb2Cu2Mo3O12_0} S. F. Solodovnikov and Z. A. Solodovnikova,  Zh. Strukt. Khim. \textbf{38} 914 (1997) 
\bibitem{Rb2Cu2Mo3O12} M. Hase, H. Kuroe, K. Ozawa, O. Suzuki, H. Kitazawa, G. Kido, and T. Sekine, Phys. Rev. B \textbf{70}, 104426 (2004).
\bibitem{PbCuSO4OH_0} H. Effenberger, Mineralogy and Petrology \textbf{36}, 3 (1987).
\bibitem{PbCuSO4OH_2} A. U. B. Wolter, F. Lipps, M. Sch\"{a}pers, S.-L. Drechsler, S. Nishimoto, R. Vogel, V. Kataev, B. Buchner, H. Rosner,
M. Schmitt, M. Uhlarz, Y. Skourski, J. Wosnitza, S. Sullow, and K. C. Rule, Phys. Rev. B \textbf{85}, 014407 (2012).
\bibitem{PbCuSO4OH_3} B. Willenberg, M. Sch\"{a}pers, K. C. Rule, S. S\"{u}llow, M. Reehuis, H. Ryll, B. Klemke, K. Kiefer, W. Schottenhamel, B. B\"{u}chner, B. Ouladdiaf, M. Uhlarz, R. Beyer, J. Wosnitza, and A. U. B. Wolter, Phys. Rev. Lett. \textbf{108}, 117202 (2012).
\bibitem{PbCuSO4OH_4} A. U. B. Wolter, F. Lipps, M. Sch\"{a}pers, S.-L. Drechsler, S. Nishimoto, R. Vogel, V. Kataev, B. Buchner, H. Rosner, M. Schmitt, M. Uhlarz, Y. Skourski, J. Wosnitza, S. Sullow, and K. C. Rule, Phys. Rev. B \textbf{85}, 014407 (2012).
\bibitem{PbCuSO4OH_6} M. Sch\"{a}pers, A. U. B. Wolter, S.-L. Drechsler, S. Nishimoto, K.-H. M\"{u}ller, M. Abdel-Hafiez, W. Schottenhamel, B. B\"{u}chner, J. Richter, B. Ouladdiaf, M. Uhlarz, R. Beyer, Y. Skourski, J. Wosnitza, K. C. Rule, H. Ryll, B. Klemke, K. Kiefer, M. Reehuis, B. Willenberg, and S. S\"{u}llow, Phys. Rev. B \textbf{88}, 184410 (2013).
\bibitem{PbCuSO4OH_7} M. Sch\"{a}pers, H. Rosner, S.-L. Drechsler, S. S\"{u}llow, R. Vogel, B. B\"{u}chner, and A. U. B. Wolter,  Phys. Rev. B \textbf{90}, 224417 (2014).
\bibitem{PbCuSO4OH_8} B. Willenberg, M. Sch\"{a}pers, A.U.B. Wolter, S.-L. Drechsler, M. Reehuis, J.-U. Hoffmann, B. B\"{u}chner, A.J. Studer, K.C. Rule, B. Ouladdiaf, S. S\"{u}llow, and S. Nishimoto, Phys. Rev. Lett. \textbf{116}, 047202 (2016).
\bibitem{PbCuSO4OH_9} K. C. Rule, B. Willenberg, M. Sch\"{a}pers, A. U. B. Wolter, B. B\"{u}chner, S.-L. Drechsler, G. Ehlers, D. A. Tennant, R. A. Mole, J. S. Gardner, S. S\"{u}llow, and S. Nishimoto, Phys. Rev. B \textbf{95}, 024430 (2017). 
\bibitem{LiCuSbO4} S. E. Dutton, M. Kumar, M. Mourigal, Z. G. Soos, J.-J. Wen, C. L. Broholm, N. H. Andersen, Q. Huang, M. Zbiri, R. Toft-Petersen, and R. J. Cava, Phys. Rev. Lett. \textbf{108}, 187206 (2012).
\bibitem{LiCuSbO4_2} H. -J. Grafe, S. Nishimoto, M. Iakovleva, E. Vavilova, L. Spillecke, A. Alfonsov, M. -I. Sturza, S. Wurmehl, H. Nojiri, H. Rosner, J. Richter, U. K. R\''{o}ßler, S.-L. Drechsler, V. Kataev and B. B\"{u}chner, Sci. Rep. \textbf{7}, 6720 (2017).
\bibitem{LiCu2O2} R. Berger, A. Meetsma, and S. van Smaalen, J. Less-Comm. Met. \textbf{175}, 119 (1991).
\bibitem{LiCu2O2_neu} T. Masuda, A. Zheludev, B. Roessli, A. Bush, M. Markina, and A. Vasiliev, Phys. Rev. B \textbf{72}, 014405 (2005).
\bibitem{LiCu2O2_3} A. A. Bush, V. N. Glazkov, M. Hagiwara, T. Kashiwagi, S. Kimura, K. Omura, L. A. Prozorova, L. E. Svistov, A. M. Vasiliev, and A. Zheludev, Phys. Rev. B \textbf{85}, 054421 (2012).
\bibitem{3IV} H. Yamaguchi, H. Miyagai, Y. Kono, S. Kittaka, T. Sakakibara, K. Iwase, T. Ono, T. Shimokawa, and Y. Hosokoshi, Phys. Rev. B \textbf{91}, 125104 (2015).
\bibitem{TeVO4} Yu. Savina, O. Bludov, V. Pashchenko, S. L. Gnatchenko, P. Lemmens, and H. Berger, Phys. Rev. B \textbf{84}, 104447 (2011). 
\bibitem{TeVO4_2} A. Sa\'ul and G. Radtke, Phys. Rev. B \textbf{89}, 104414 (2014). 
\bibitem{TeVO4_3} M. Pregelj, A. Zorko, O. Zaharko, H. Nojiri, H. Berger, L.C. Chapon and D. Arcon, Nat. Commun. \textbf{6}, 7255 (2015). 
\bibitem{TeVO4_4} F. Weickert, N. Harrison, B. L. Scott, M. Jaime, A. Leitm\"{a}e, I. Heinmaa, R. Stern, O. Janson, H. Berger, H. Rosner, and A. A. Tsirlin, Phys. Rev. B \textbf{94}, 064403 (2016).
\bibitem{NaCuMoO4OH} K. Nawa, Y. Okamoto, A. Matsuo, K. Kindo, Y. Kitahara, S. Yoshida, S. Ikeda, S. Hara, T. Sakurai, S. Okubo, H. Ohta, and Z. Hiroi, J. Phys. Soc. Jpn. \textbf{83}, 103702 (2014).
\bibitem{NaCuMoO4OH2} A. Moini, R. Peascoe, P. R. Rudolf, and A. Clearfield, Inorg. Chem. \textbf{52}, 3782 (1986).
\bibitem{NaCuMoO4OH3} K. Nawa, Y. Okamoto, and Z. Hiroi, J. Phys. Conf. Ser. \textbf{828}, 012005 (2017).
\bibitem{NaCuMoO4OH4} S. Asai, T. Oyama, M. Soda, K. Rule, K. Nawa, Z. Hiroi and T. Masuda, J. Phys. Conf. Ser. \textbf{828}, 012006 (2017).
\bibitem{relaxation} E. R. Andrew and D. P. Tunstall, Proc. Phys. Soc. \textbf{78}, 1 (1961). 
\bibitem{relaxation2} A. Suter, M. Mali, J. Roos and D. Brinkmann, J. Phys.: Condens. Matter. \textbf{10}, 5977 (1998).
\bibitem{DFT} S. Lebernegg, A. A. Tsirlin, O. Janson, and H. Rosner, Phys. Rev. B \textbf{88}, 224406 (2013).
\bibitem{angle} Several mistakes in a theoretical function in G. C. Carter, L. H. Bennett, and D. J. Kahan,  in \textit{Progress in Materials Science}, edited by B. Chalmers et al., Vol. 20 (Pergamon Press, Oxford 1977), p. 64 are corrected.
\bibitem{demag} J. A. Osborn: Phys. Rev. \textbf{67}, 351 (1945).
\bibitem{hyperfine2} J. Owen and J. H. M. Thornley, Rep. Prog. Phys. \textbf{29}, 675 (1966).
\bibitem{hyperfine} M. W. van Tol, K. M. Diederix and N. J. Poulis, Physica \textbf{64}, 363 (1973) (Commun. Kamerlingh Onnes Lab., Leiden No. 397c).
\bibitem{deconvolution} F. Mila and M. Takigawa, Eur. Phys. J. B \textbf{86}, 354 (2013).
\bibitem{NMR6} The temperature dependence of the second moment is determined from NMR spectra reported in Ref.~\onlinecite{NMR2}.
\bibitem{1DtheoryofT11} M. Sato, T. Momoi, and A. Furusaki, Phys. Rev. B \textbf{79}, 060406 (2009).
\bibitem{1DtheoryofT12} M. Sato, T. Hikihara, and T. Momoi, Phys. Rev. B \textbf{83}, 064405 (2011).
\end{thebibliography}
\end{document}